\documentclass[preprint]{aastex61}

\usepackage{amsmath}
\usepackage{amssymb}
\usepackage{graphicx}
\usepackage{bm}
\usepackage{geometry}
\usepackage{natbib}
\usepackage{extarrows}
\usepackage{color}
\usepackage{ulem}

\begin{document}

\title{reanalyze the pulsar scenario to explain the cosmic positron excess
considering the recent developments}

\author{Kun Fang}
\affiliation{Key Laboratory of Particle Astrophysics, Institute of High Energy 
Physics, Chinese Academy of Sciences, Beijing 100049, China}

\author{Xiao-Jun Bi}
\affiliation{Key Laboratory of Particle Astrophysics, Institute of High Energy 
Physics, Chinese Academy of Sciences, Beijing 100049, China}
\affiliation{School of Physics, University of Chinese Academy of Sciences, 
Beijing 100049, China} 

\author{Peng-Fei Yin}
\affiliation{Key Laboratory of Particle Astrophysics, Institute of High Energy 
Physics, Chinese Academy of Sciences, Beijing 100049, China}

\date{\today}

\begin{abstract}
As the TeV halos around Geminga and PSR B0656+14 have been confirmed by HAWC, 
slow diffusion of cosmic rays could be general around pulsars, and the cosmic 
positron spectrum from pulsars could be significantly changed. As 
a consequence, the most likely pulsar source of the positron excess, Geminga, is 
no more a viable candidate under the additional constraint from Fermi-LAT. 
Moreover, the latest measurement by AMS-02 shows a clear cutoff in the positron 
spectrum, which sets a strict constraint on the age of the pulsar source. 
Considering these new developments we reanalyze the scenario in this work. By 
checking all the observed pulsars under the two-zone diffusion scenario, we 
propose for the first time that PSR B1055-52 is a very promising source of the 
positron excess. B1055-52 can well reproduce both the intensity and the 
high-energy cut of the AMS-02 positron spectrum, and may also explain the 
H.E.S.S $e^-+e^+$ spectrum around 10 TeV. Moreover, if the slow diffusion is 
universal in the local interstellar medium, B1055-52 will be the unique 
reasonable source of the AMS-02 positron spectrum among the observed pulsars.
\end{abstract}

\section{Introduction}
\label{sec:intro}
The well-known cosmic-ray (CR) positron excess, which was first detected by 
PAMELA \citep{2009Natur.458..607A} and confirmed by AMS-02 
\citep{2013PhRvL.110n1102A} and Fermi-LAT \citep{2012PhRvL.108a1103A} later, may 
be interpreted by either astrophysical sources or the dark matter 
annihilation/decay. The dark matter scenarios are strongly disfavored by the 
$\gamma$-ray observation of Fermi-LAT on the dwarf galaxies and so on 
\citep{2015PhRvL.115w1301A,2017ChPhC..41d5104L,2015PhRvD..91f3508L}. Nearby 
pulsars, such as Geminga, PSR B0656+14, and Vela pulsar, have been proposed
as plausible sources of high-energy positrons as buttressed by quantitative 
calculations \citep{2009JCAP...01..025H,2017PhRvD..96j3013H,2009PhRvL.103e1101Y,
2013PhRvD..88b3001Y,2015JHEAp...8...27D}. It is also possible that the excess is 
contributed by multiple nearby pulsars \citep{dela10}.

The study of the positron excess is now entering into a new stage as recent
developments change the situation rapidly. First, the TeV $\gamma$-ray 
observation of HAWC indicates that Geminga and B0656+14 are surrounded by very 
inefficient diffusion halos \citep{2017Sci...358..911A}, and it is possible 
that the slow-diffusion region is general around pulsars 
\citep{2017PhRvD..96j3016L}. If so, the positron contribution from pulsars 
should be significantly different from that derived by the previous one-zone 
propagation model. Second, the AMS collaboration has just published the latest 
positron spectrum using 6.5 
years of data \citep{2019PhRvL.122d1102A}. The positron spectrum is 
unprecedentedly extended to $\sim$ 1 TeV, and a sharp spectral dropoff around 
300 GeV is detected for the first time with a significance of more than 
$3\sigma$. This definitely provides a stronger limit to the high-energy 
positron source.

In \citet{2018ApJ...863...30F}, we perform the first numerical calculation of 
the positron spectrum of pulsars at the Earth under the two-zone diffusion 
scenario, in which the diffusion velocity outside the slow-diffusion halo 
around the pulsars is assumed to be the average value in the Galaxy. In this 
case, Geminga can naturally explain the positron excess if we take the 
injection spectral index derived by HAWC \citep{2017Sci...358..911A}. However, 
analysis of the Fermi-LAT data on the Geminga halo shows that the injection 
spectrum in GeV should be much harder than that provided by HAWC 
\citep{2018arXiv181010928X,2019arXiv190305647D}. As a result, the conversion 
efficiency to positrons is significantly depressed, and Geminga is disfavored 
as the dominant source of the high-energy positrons 
\citep{2018arXiv181010928X,2019arXiv190305647D}.

In this work, we give a reanalysis of the pulsar candidates to explain the 
positron excess considering the new experimental developments. We assume the 
slow-diffusion halo is universal around pulsars and adopt the two-zone 
diffusion model to all the observed nearby pulsars to calculate their positron 
fluxes at the Earth. We consider that the positron spectral cutoff detected by 
AMS-02 should not be a collective effect of pulsars, since even old pulsar wind 
nebulae (PWNe) can accelerate positrons to $\sim100$ TeV 
\citep{2003Sci...301.1345C,2017Sci...358..911A}. The spectral break in hundreds 
of GeV is more likely to correspond to the radiative cooling of the positrons 
released by the dominant source, which will set a strict constraint to the 
pulsar age. We attempt to find the promising pulsar(s) that can both explain 
the intensity and the spectral cut of the the latest positron spectrum of 
AMS-02.

In the next section, we describe the details of the calculation of the positron 
spectrum from pulsars. In Section \ref{sec:result}, we present the prominent 
sources of positrons among the observed pulsars under the two-zone 
diffusion model, and also discuss the impacts from the uncertainties of the 
model. In Section \ref{sec:discuss}, we give further discussions about the most 
prominent positron source found in Section \ref{sec:result}: PSR B1055-52. We 
summarize this work in Section \ref{sec:sum}.

\section{Method}
\label{sec:method}

\subsection{The two-zone propagation model}
\label{subsec:prop}
The propagation of CR electrons and positrons ($e^\pm$) is generally described 
by the diffusion-cooling equation:
 \begin{equation}
  \frac{\partial N}{\partial t} - \nabla\cdot(D\nabla N) - 
\frac{\partial}{\partial
 E}(bN) = Q \,,
  \label{eq:prop}
 \end{equation}
where $D$ is the diffusion coefficient, $b(E)=-{\rm d}E/{\rm d}t$ is the 
energy-loss rate, and $Q$ denotes the source function. The energy-loss rate 
takes the form of $b(E)=b_0(E)E^2$, and we calculate $b_0(E)$ following 
\citet{schli10}, where the synchrotron and inverse Compton radiation cooling of 
$e^\pm$ are included. We assume the interstellar magnetic field in the Galaxy 
to be 3 $\mu$G as the default to calculate the synchrotron term 
\citep{1996ApJ...458..194M}.

As the diffusion coefficient derived by HAWC is hundreds times slower than 
the Galactic average value \citep{2016PhRvL.117w1102A}, this slow-diffusion 
velocity in the vicinity of the pulsars should not be universal in the Galaxy 
\citep{2017PhRvD..96j3013H}. The slow-diffusion halo\footnote{Recently,
\citet{2019arXiv190411536L} propose an alternative scenario that the TeV halo 
of Geminga is not attributed to the strong turbulence, but interpreted by the
anisotropy diffusion of the electrons along the local regular magnetic field 
which is presumed to be aligned with the line of sight towards Geminga.} could 
either be self-induced by the escaping $e^\pm$ from the pulsar 
\citep{2018PhRvD..98f3017E}, or simply be a preexisting turbulent region 
created by the parent supernova remnant of the pulsar 
\citep{2019arXiv190306421F}. Thus, the slow-diffusion regions may have a scale 
of tens of pc considering their possible origins. We write the diffusion 
coefficient as
\begin{equation}
 D(E, r)=\left\{
 \begin{aligned}
  D_1(E), & & r< r_\star \\
  D_2(E), & & r\geq r_\star\\
 \end{aligned}
 \right.\,,
 \label{eq:diff}
\end{equation}
where $r$ is the distance to the pulsar, $D_1(E)=10^{26}(E/1 {\rm 
\,GeV})^{0.33}$ cm$^{2}$ s$^{-1}$ is the diffusion coefficient inferred by 
HAWC, $D_2(E)=4.28\times10^{28}(E/1 {\rm\,GeV})^{0.38}$ cm$^{2}$ s$^{-1}$ is 
the average diffusion coefficient in the Galaxy given by the 
diffusion-reacceleration model of \citet{2017PhRvD..95h3007Y}, and we assume a 
spherically symmetrical diffusion for the $e^\pm$ released by local pulsars. 
The size of the slow-diffusion region is set to be 50 pc as the default. 

For variable coefficient problems like Equation (\ref{eq:prop}) and 
(\ref{eq:diff}), numerical method should be the more straightforward choice. We 
follow the method given in \citet{2018ApJ...863...30F}. The finite volume 
method is adopted to construct the differencing scheme for the diffusion 
operator. We emphasize that this is necessary to ensure the conservation of 
flux for the  variable-coefficient diffusion equation and obtain the correct 
solution.

\subsection{Pulsars as the positron source}
We assume pulsar as point-like source with continuous $e^\pm$ injection, so 
the source function should be expressed as
\begin{equation}
 Q(t,E,r)=q(t,E)\delta(r-r_s)\,,
 \label{eq:source}
\end{equation}
where $r_s$ is the distance of the pulsar. The time dependency of $e^\pm$ 
injection is assumed to have the same profile with the spin-down luminosity of 
the pulsar \citep{1973ApJ...186..249P}, then
\begin{equation}
 q(t,E)=q_0\left(1+\frac{t}{\tau}\right)^{-2}E^{-\gamma}\,,
 \label{eq:time_profile}
\end{equation}
where $\tau=10$ kyr is the typical the spin-down time scale of the pulsar. Note 
$\gamma$ is the injection spectral index of the PWN, rather than that of the 
pulsar. Radio observations indicate that the $e^\pm$ spectral index is smaller 
than 2.0 for most PWNe \citep{2017SSRv..207..175R}. Besides, the injection 
spectral index of Geminga also tends to be hard considering the constraint from 
the observation of Fermi-LAT \citep{2018arXiv181010928X,2019arXiv190305647D}. 
So 
we set $\gamma=1.8$ as the default in the following calculation. The 
normalization $q_0$ is determined by the relation
\begin{equation}
 \int_{E_{\rm min}}^{E_{\rm max}}q(t_s,E)EdE=\eta\dot{E}_s\,,
 \label{eq:norm}
\end{equation}
where we set $E_{\rm min}=1$ GeV, $E_{\rm max}=100$ TeV, $t_s$ and $E_s$ are 
the current age and spin-down luminosity of the pulsar, respectively. Equation 
(\ref{eq:norm}) means the spin-down energy of the pulsar is converted to the 
injected $e^\pm$ with a proportion of $\eta$.

The $r_s$, $t_s$, and $E_s$ of all the published pulsars can be found in the 
ATNF catalog\footnote{http://www.atnf.csiro.au/research/pulsar/psrcat} 
\citep{2005AJ....129.1993M}. For the measurement of 
the pulsar distance,  the kinematic method and the trigonometric parallax take 
the precedences in the ATNF catalog. For the cases in which those methods are 
unavailable, the pulsar distance is derived from the dispersion measure (DM), 
which depends on a specific electron distribution model. We should note that 
for the version later than 1.55 (Nov 2016), the electron distribution model of 
\citet{2017ApJ...835...29Y} (YMW16) is adopted as the default instead of those 
of \citet{1993ApJ...411..674T} (TC93) and \citet{2002astro.ph..7156C} (NE2001). 
The distance estimates of some pulsars are significantly changed under the 
YMW16 model. For example, the distance of PSR J0940-5428 is 0.38 kpc using the 
YMW16 model, compared with 2.95 kpc based on the NE2001 model; the distance of 
PSR B1055-52 is 0.09 kpc using the YMW16 model, compared with 0.72 kpc based on 
the NE2001 model.

\begin{figure}[t]
 \centering
 \includegraphics[width=0.7\textwidth]{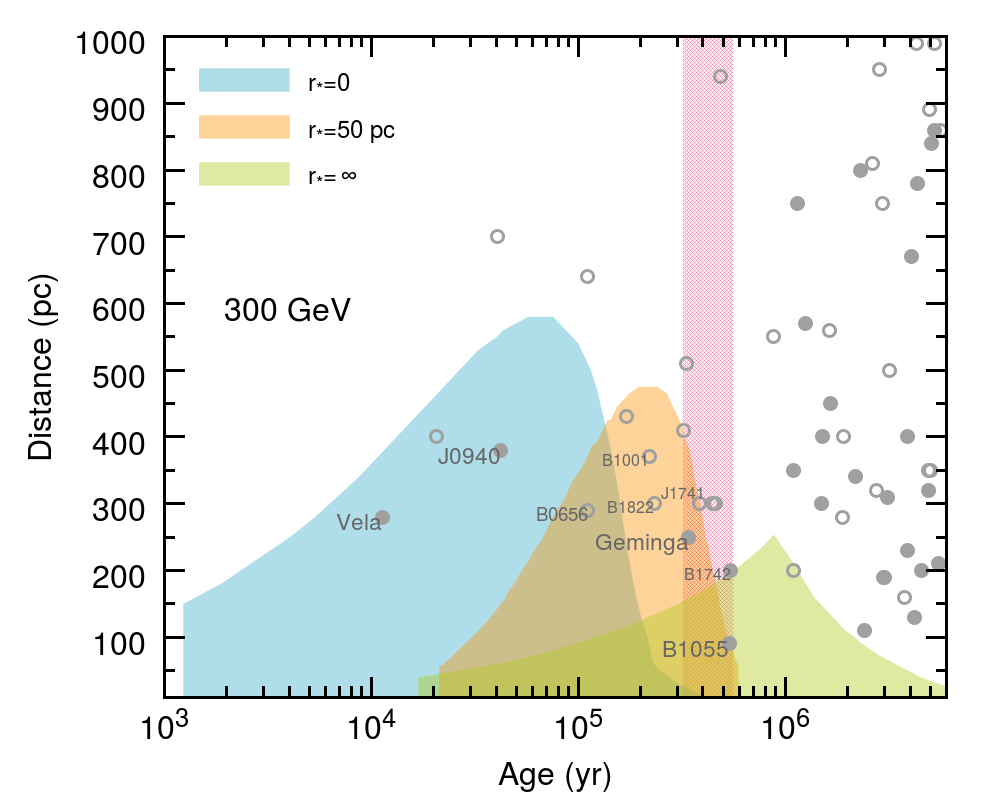}
 \caption{Contours of the positron flux at 300 GeV as a function of age and 
distance of pulsar. We assume $\gamma=1.8$, $\eta=100\%$, and a 
log-mean $\dot{E}_{0,s}$ of the observed pulsars. The shaded areas means that 
the pulsars inside this area can contribute more than half of the measured flux 
of AMS-02 at 300 GeV, and different colors represent different diffusion 
models. Observed pulsars are marked with circles. Filled ones are those with 
$\dot{E}_{0,s}$ larger than the log-mean value, while empty ones are on the 
contrary. The red band shows the age range corresponding to positron spectral 
cut.}
 \label{fig:contour_300gev}
\end{figure}

\section{Result}
\label{sec:result}
As mentioned above, we set $\gamma=1.8$ and $r_\star=50$ pc as the defaults to 
calculate the positron spectrum of pulsars at the Earth. Figure 
\ref{fig:contour_300gev} shows the contours of the positron flux at 300 GeV as 
a function of age and distance of pulsars. The yellow shaded area means that 
the pulsars inside this area can contribute more than half of the measured 
positron flux of AMS-02 at 300 GeV under the two-zone diffusion model. For 
comparison, we also draw the blue and green areas corresponding to the one-zone 
fast-diffusion model ($r_\star=0$) and the one-zone slow-diffusion model 
($r_\star=\infty$), respectively. Any two-zone diffusion scenarios should be 
the transitions among these three models. In the calculation, we use a 
uniform initial spin-down luminosity $\dot{E}_{0,s}$ to determine the 
normalization $q_0$, and the log-mean $\dot{E}_{0,s}$ of all the nearby ($<1$ 
kpc) and relatively young ($<10$ Myr) pulsars are adopted. We assume all the 
spin-down energy is delivered to $e^\pm$ ($\eta=100\%$). The observed pulsars 
in the ATNF catalog are shown in Figure \ref{fig:contour_300gev}. The pulsars 
of which the $\dot{E}_{0,s}$ is larger than the log-mean value mentioned above 
are marked with filled circles, otherwise they will be marked with unfilled 
ones. Thus, the upper limits of the shaded areas should be in fact somehow 
higher than those shown in the figure for the filled circles, and lower for the 
unfilled circles.

\citet{2019PhRvL.122d1102A} fit the AMS-02 positron spectrum with a diffuse 
term and a high energy source term, and the cutoff energy of the source term 
is $E_c=810^{+310}_{-180}$ GeV. If the cutoff is interpreted by the cooling of 
the positrons released in the early age of a pulsar, we may approximately 
relate the age of the pulsar to the cooling time $1/[b_0(E_c)E_c]$. In Figure 
\ref{fig:contour_300gev}, we show this age range with the red band for 
reference.

\begin{table}[t]
\centering
 \begin{tabular}{ccccc}
  \hline
  \hline
  J2000 & Other Name & $t_s$ & $r_s$ & $\dot{E}_s$ \\
   & & (kyr) & (kpc) & ($10^{34}$ erg s$^{-1}$) \\
  \hline
  J0633+1746 & Geminga & 342 & 0.25 & 3.25 \\
  \hline
  J0659+1414 & B0656+14 & 111 & 0.29 & 3.81 \\
  \hline
  J1003-4747 & B1001-47 & 220 & 0.37 & 3.01 \\
  \hline
  J1057-5226 & B1055-52 & 535 & 0.09 & 3.01 \\
  \hline
  J1741-2054 & - & 386 & 0.30 & 0.95 \\
  \hline
  J1745-3040 & B1742-30 & 546 & 0.20 & 0.85 \\
  \hline
  J1825-0935 & B1822-09 & 232 & 0.30 & 0.46 \\
  \hline
 \end{tabular}
 \caption{Candidate pulsars of the positron excess under the two-zone 
diffusion model.}
\label{tab:pulsars}
\end{table}

\subsection{Candidate sources of the positron excess}
\begin{figure}[t]
 \centering
 \includegraphics[width=0.48\textwidth]{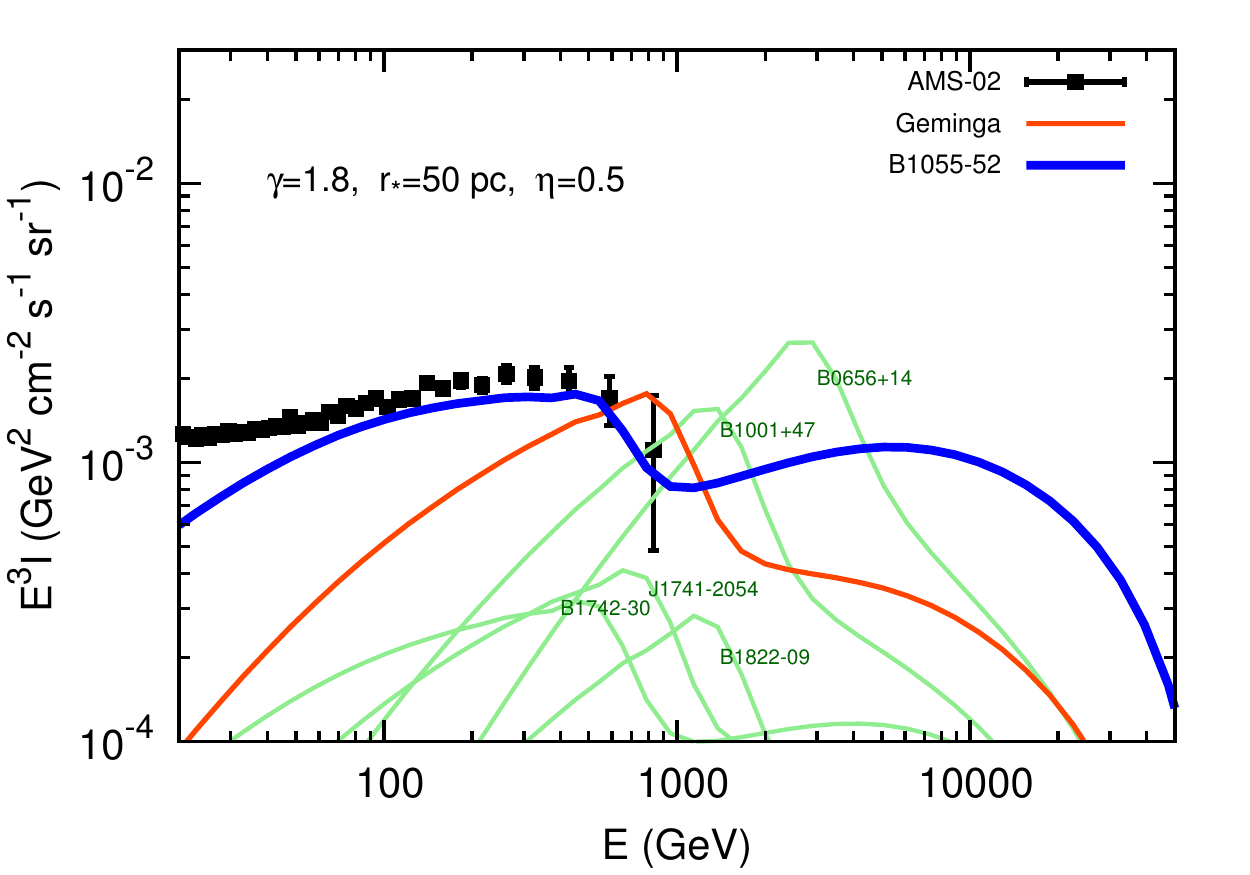}
 \includegraphics[width=0.48\textwidth]{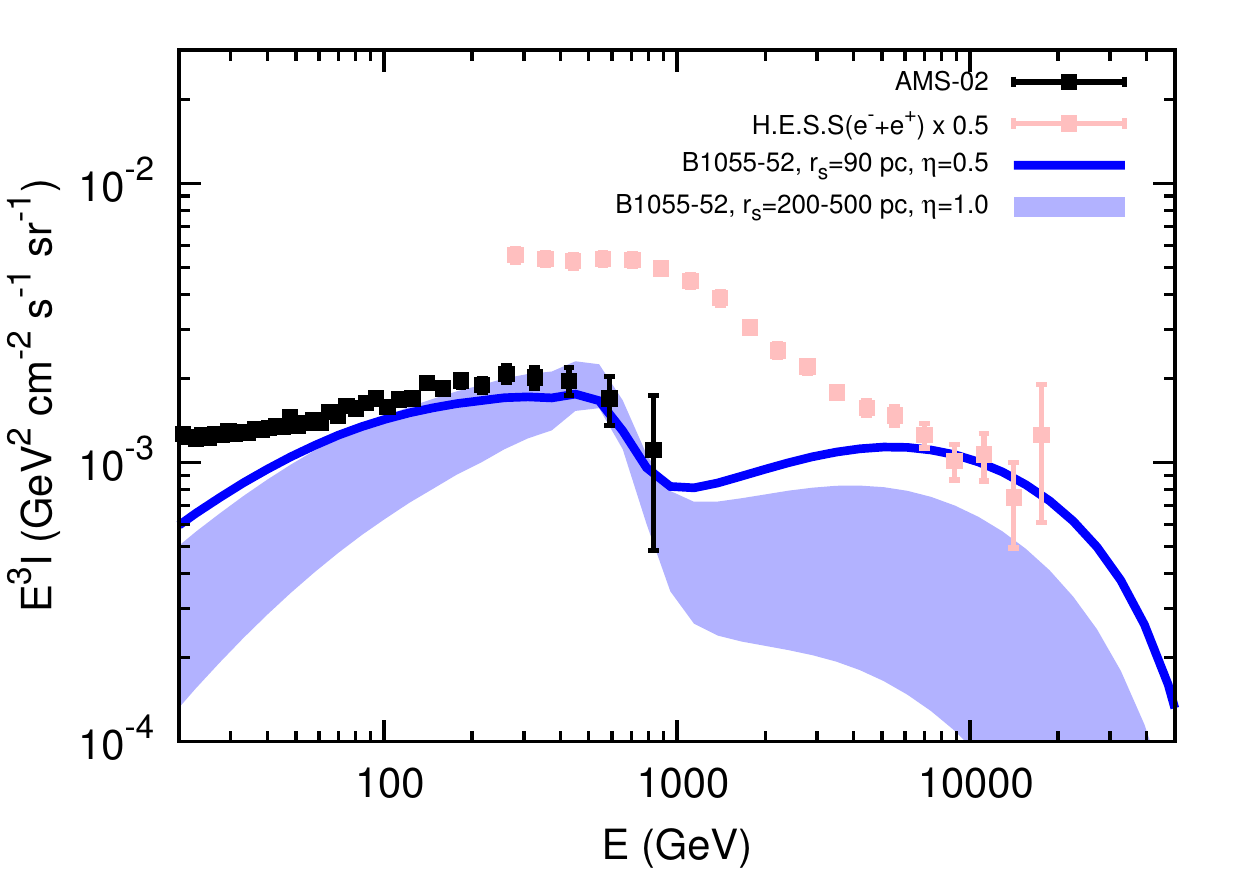}
 \caption{Left: positron spectra from the nearby and middle-aged pulsars under 
the two-zone diffusion model, compared with the latest AMS-02 data. Right: 
positron spectra from B1055-52 with the different distances given by DM (line) 
and impied by \citet{2010ApJ...720.1635M} (band). The $e^-+e^+$ spectrum of 
H.E.S.S with half of the flux is also shown.}
 \label{fig:2zone_50_1d8}
\end{figure}

Figure \ref{fig:contour_300gev} indicates that in the two-zone diffusion 
scenario the positron excess can only be contributed by nearby ($r_s<500$ pc) 
and middle-aged ($t_s\sim0.1-1$ Myr) pulsars. As $e^\pm$ injected by pulsars 
are trapped in the slow-diffusion region for a long time, $e^\pm$ released 
by young pulsars do not have enough time to arrive at the Earth. Table 
\ref{tab:pulsars} summarizes the relevant information of the pulsars inside the 
yellow area of Figure \ref{fig:contour_300gev}. We show the positron spectrum 
of these pulsars at the Earth in Figure \ref{fig:2zone_50_1d8}, with 
$\eta=50\%$ for all the pulsars. Obviously, B1055-52 and Geminga are the most 
prominent positron sources in the energy range of AMS-02 with our default 
parameters. Moreover, B1055-52 can well reproduce the cutoff of the AMS-02 
spectrum, while Geminga is not old enough to fit this spectral fall.

B1055-52, together with Geminga and B0656+14, are called \textit{the three 
musketeers} as they have similar ages, spin-down luminosities, and so on 
\citep{1997A&A...326..682B}. All of them are $\gamma$-ray pulsars 
\citep{2010ApJS..187..460A}, and surrounded by X-ray PWNe 
\citep{2003Sci...301.1345C,2015ApJ...811...96P,2016ApJ...817..129B,
2017ApJ...835...66P}. However, B1055-52 has rarely been discussed when it 
comes to the interpretation of the positron excess, because it was once 
believed to be located much farther according to the previous models of the
Galactic electron distribution (1.53 kpc for TC93 and 720 pc for NE2001). The 
very close distance of 90 pc indicated by the YMW16 model ensures a sufficient 
positron contribution to the AMS-02 data. High-energy positrons injected at the
later age of B1055-52 also have enough time to reach the Earth, which may even 
explain the highest energy part of the H.E.S.S spectrum\footnote{
https://indico.snu.ac.kr/indico/event/15/session/5/contribution/694} as shown 
in the right of Figure \ref{fig:2zone_50_1d8}. Meanwhile, 
\citet{2010ApJ...720.1635M} also argued that B1055-52 should be a nearby 
source, but with a distance of $350\pm150$ pc which is farther than that given 
by the latest DM. In this case, a higher conversion efficiency for B1055-52 is 
required to explain the positron spectrum. In the right of Figure 
\ref{fig:2zone_50_1d8}, we show the case of $r_s=350\pm150$ pc and $\eta=100\%$ 
for B1055-52 with the blue band.

Geminga has been widely discussed as the source of the positron excess 
\citep{2009JCAP...01..025H,2017PhRvD..96j3013H,2009PhRvL.103e1101Y,
2013PhRvD..88b3001Y,2018ApJ...863...30F,2018PhRvD..97l3008P,2019MNRAS.484.3491T}
. In \citet{2018ApJ...863...30F}, we point out that Geminga can reasonably 
explain the positron spectrum of AMS-02 in the two-zone diffusion scenario, 
with $\gamma=2.2$ provided by \citet{2017Sci...358..911A}. However, 
$\gamma$-ray measurement of Fermi-LAT around Geminga implies that the $e^\pm$ 
spectrum cannot be extrapolated to GeV energies with $\gamma=2.2$. The spectrum 
should be harder and the conversion efficiency is constrained to be less than 
$5\%$, which disfavors Geminga as the main source of positron excess 
\citep{2018arXiv181010928X}. While, the $e^\pm$ accounting for the AMS-02 
spectrum should be mainly released in the early age of Geminga. Considering the 
proper motion of Geminga and the uncertainties of the diffusion pattern in the 
early time, Geminga may not be finally excluded as the positron source.

\begin{figure}[t]
 \centering
 \includegraphics[width=0.48\textwidth]{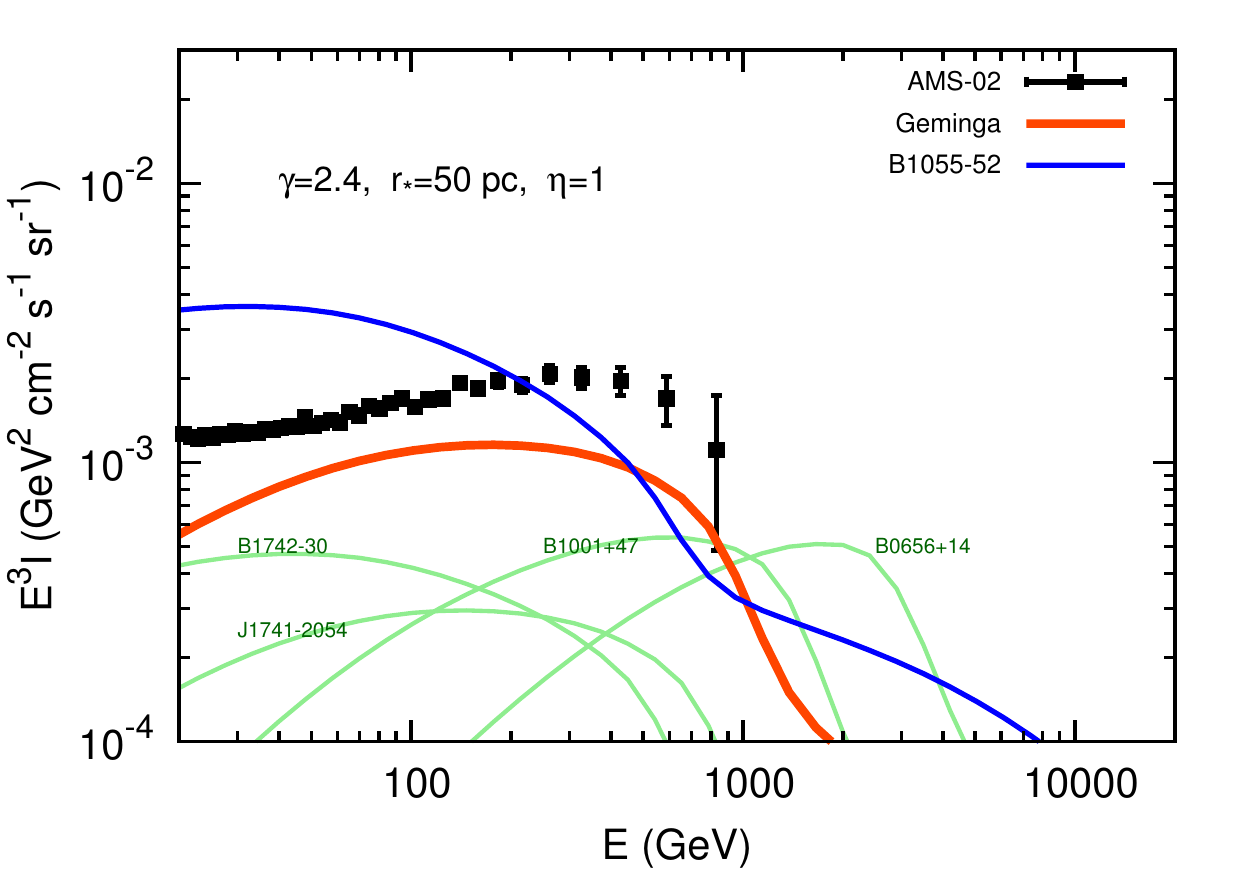}
 \includegraphics[width=0.48\textwidth]{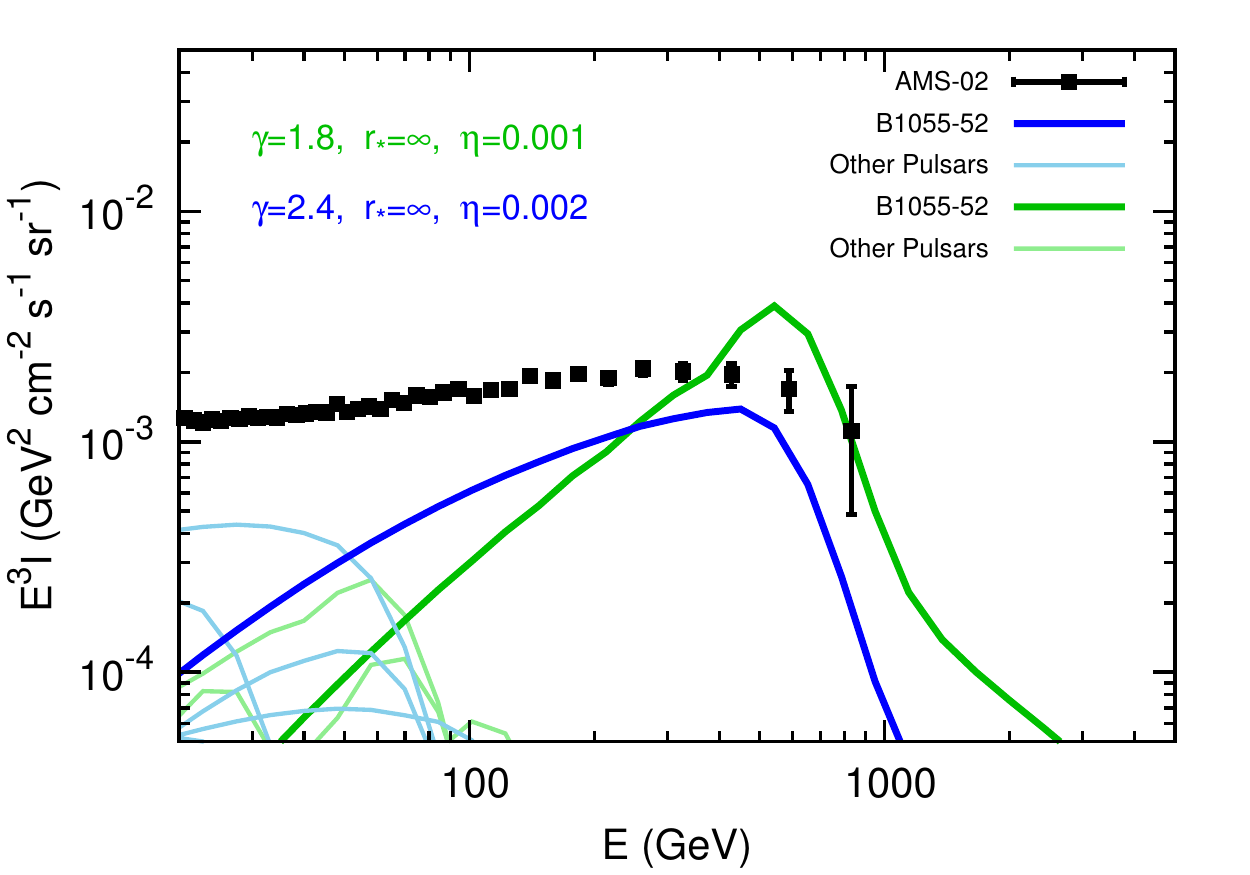}
 \caption{Left: same as the left panel of Figure \ref{fig:2zone_50_1d8}, but 
with a much softer injection spectral index of $\gamma=2.4$. Right: the 
scenario in which the slow diffusion is universal in the local interstellar 
medium.}
 \label{fig:uncerain}
\end{figure}

\subsection{Uncertainties of the model}
Here we discuss the main uncertainties in the calculation above: $\gamma$, 
$r_\star$, and $B$. Observations favor a hard injection $e^\pm$ spectrum for 
PWNe, while if the $e^\pm$ are accelerated at relativistic shocks (the 
termination shocks), theories predict a softer spectrum with $\gamma\geq2.2$ 
\citep{1999JPhG...25R.163K}. So we present a case with $\gamma=2.4$ in the left 
of Figure \ref{fig:uncerain}, where the injection spectrum is much softer than 
the default. B1055-52 provides too much low-energy positrons in this 
case, while we should note that there is degeneracy between $\gamma$ and 
$r_\star$ in terms of the positron spectrum at the Earth. So if $r_\star$ is 
larger, the spectral shape of B1055-52 may still match the measurement. The 
spectral shape of Geminga has a better consistency with the AMS-02 data for 
$\gamma=2.4$, since the spectral cut due to radiative cooling appears at lower 
energy compared with the case of $\gamma=1.8$. However, due to the constraint 
from Fermi-LAT, the scenario of $\gamma>2.0$ for Geminga is excluded 
\citep{2018arXiv181010928X}.

The size of the slow-diffusion halo around pulsars cannot be generally 
constrained by observations at present. More TeV halos associated with pulsars 
need to be confirmed. What is certain is that we have $r_\star\geq20$ pc around 
Geminga and B0656+14 \citep{2017Sci...358..911A}. Besides, if the extend TeV 
emission around PSR J1826-1334 and J1907+0602 (HESS J1825-137 and MGRO J1908+06 
respectively) are attributed to the inefficient diffusion of $e^\pm$, the scale 
of $r_\star$ should be at least tens of pc 
\citep{2018ApJ...860...59K,2014ApJ...787..166A}. \citet{2018ApJ...866..143H} 
also argue that a considerable slow diffusion should be existed around the Vela 
pulsar to avoid the conflict with the high-energy $e^-+e^+$ spectrum of 
H.E.S.S. Moreover, Figure \ref{fig:contour_300gev} indicates that if $r_\star$ 
is null or too small, pulsars cannot explain the positron flux and the positron 
spectral cut of AMS-02 simultaneously (for example, the blue shaded area has no 
overlap with the red area). All these imply that the size of the 
slow-diffusion region could be considerable for pulsars. However, the upper 
limit of $r_\star$ is more difficult to be decided. In the right of Figure 
\ref{fig:uncerain}, we show the most extreme scenario with $r_\star=\infty$, 
that is, the slow diffusion is universal in the local interstellar medium 
(ISM). In this case, only very nearby source could explain the high-energy 
positron spectrum. Obviously, B1055-52 is the best candidate as it is the 
nearest pulsar in the ATNF catalog and it also has a proper age. Meanwhile, 
only a very small conversion efficiency of $0.1\%-0.2\%$ is required to explain 
the data.

We have assumed a typical interstellar magnetic field of 3 $\mu$G in the 
calculations above. \citet{2018MNRAS.479.4526L} argue that the local magnetic 
field may be in a range of $3-5$ $\mu$G. If $B>3$ $\mu$G, the cooling cut of 
pulsars in the positron spectrum should be happened at lower energies compared 
with Figure \ref{fig:2zone_50_1d8} and \ref{fig:uncerain}, which means the 
spectral cut of younger pulsars such as B1001+47 may match the cut of AMS-02. 
However, the spin-down luminosity of these observed younger sources are too low 
to generate enough positrons. So the conclusion is unchanged and B1055-52 
should still be the most likely candidate.

\section{Discussion}
\label{sec:discuss}

\subsection{The distance of PSR B1055-52}
\label{subsec:b1055_dist}
We have shown that B1055-52 is a very competitive source of the positron excess 
under the two-zone diffusion scenario. An important factor of its dominance is 
the very close distance of 90 pc. This distance is given by the DM based on the 
electron density model of YMW16, while the annual parallax of this source has 
not been measured yet. 

The relation of the $\gamma$-ray luminosity $L_\gamma$ and $\dot{E}_s$ could be 
a supporting evidence for the distance of B1055-52. \citet{2010ApJS..187..460A} 
show the $L_\gamma-\dot{E}_s$ map for the observed $\gamma$-ray pulsars. Most 
pulsars with high $\dot{E}_s$ are distributed between $L_\gamma=\dot{E}_s$ and 
$L_\gamma=(10^{33}\,{\rm erg s}^{-1}\dot{E}_s)^{1/2}$, while there seems to be 
a 
break in $\dot{E}_s\sim10^{35}$ erg s$^{-1}$, that is, the $\gamma$-ray 
efficiency drops faster with the decrease of $\dot{E}_s$. This is consistent 
with the theoretical predictions 
\citep{2003ApJ...588..430M,2004ApJ...604..317Z}. The $\gamma$-ray efficiency of 
B0656+14 is very low, which has $L_\gamma=3.1\times10^{32}$ erg s$^{-1}$ and 
$\dot{E}_s=3.8\times10^{34}$ erg s$^{-1}$. If the distance of B1055-52 is 90 
pc, then B1055-52 is almost coincident with B0656+14 on the 
$L_\gamma-\dot{E}_s$ map, with $L_\gamma=2.6\times10^{32}$ erg s$^{-1}$ and 
$\dot{E}_s=3.0\times10^{34}$ erg s$^{-1}$. So we consider $r_s=90$ pc is 
reasonable for B1055-52. First, B1055-52 is similar with B0656+14 in most 
aspects. Their X-ray PWNe are both very faint or compact 
\citep{2015ApJ...811...96P,2016ApJ...817..129B}, which is different from that 
of Geminga \citep{2017ApJ...835...66P}. Second, the distance of B0656+14 is 
given by the method of trigonometric parallax with very small uncertainty 
\citep{2003ApJ...593L..89B}, which means the position of B0656+14 on the 
$L_\gamma-\dot{E}_s$ map is exact.

\subsection{Could the TeV halo of B1055-52 be detected?}
\label{subsec:b1055_halo}
TeV halos around pulsars are strong evidence of slow diffusion. Like Geminga 
and B0656+14, B1055-52 should also own a TeV $\gamma$-ray halo if it is 
surrounded by a slow-diffusion region. However, the location of B1055-52 is 
($286.0^\circ$, $+6.6^\circ$) in the Galactic coordinate and ($164.5^\circ$, 
$-52.5^\circ$) in the equatorial coordinate, which is out of the sight of HAWC 
\citep{2017ApJ...843...40A}. H.E.S.S has performed Galactic plane survey with 
high quality data recored from 2004 to 2013 \citep{2018A&A...612A...1H}, while 
the survey only covers the Galactic latitude from $-3^\circ$ to $3^\circ$, and 
the area of B1055-52 is still not included in the survey. Besides, if the 
radius of the TeV halo is 20 pc, the field angle of the halo should be 
$25^\circ$ for $r_s=90$ pc, which is too large for the imaging atmospheric 
Cherenkov telescopes including H.E.S.S. Thus, a large field-of-view TeV 
$\gamma$-ray observatory in the southern hemisphere with large field of view, 
such as the SGSO which is proposed recently \citep{2019arXiv190208429A}, is 
necessary to search for the TeV halo around B1055-52. 


\section{Conclusion}
\label{sec:sum}
As the slow-diffusion regions around Geminga and B0656+14 have been confirmed 
by HAWC, slow diffusion around pulsars could be general. 
In this work, we interpret the latest AMS-02 positron spectrum with identified 
pulsars under the two-zone diffusion scenario. The intensity and the 
high-energy spectral cut of the AMS-02 spectrum are required to be explained 
simultaneously. We consider that the spectral drop is attributed to the 
cooling of the positrons from the dominate pulsar.

We provide the contours of the positron contribution from pulsar as a function 
of age and distance of pulsar. Three different diffusion cases are 
shown, with $r_\star=0$, $r_\star=50$ pc, and $r_\star=\infty$. Any two-zone 
diffusion scenarios should be the transitions among these models. The 
contours clearly show that if the scale of the slow-diffusion halo is 
significant ($\sim50$ pc), the high-energy positron spectrum tends to be 
explained by \textit{nearby} and \textit{middle-aged} pulsars. Meanwhile, 
$r_\star$ should not be generally null or too small, otherwise no pulsar is 
able to reproduce the intensity and the spectral cut of the AMS-02 data at the 
same time.

We scan all the observed pulsars, and find that B1055-52 and Geminga are the 
most promising candidates of the positron excess, although Geminga is 
disfavored by the constraint of the $\gamma$-ray observation of Fermi-LAT. 
B1055-52, which has the similar age and spin-down luminosity with Geminga and 
B0656+14, was rarely discussed in terms of the positron excess, because it 
was too far as indicated by the previous DM. The latest DM derives a very close 
distance of 90 pc for B1055-52, and it becomes the brightest positron source, 
which may even explain the H.E.S.S $e^-+e^+$ spectrum around 10 TeV. Besides, 
even if the slow diffusion is universe in the local ISM, B1055-52 is still able 
to explain the positron excess and it is the only candidate in this scenario. 

We also discuss that the distance of 90 pc can be reasonable for B1055-52, 
while we suggest the annual parallax measurement for B1055-52 to obtain a more 
precise distance. The putative TeV halo around B1055-52 cannot be detected by 
the current experiments, due to the limit of their locations or the field of 
view. Future wide field-of-view TeV observatory in the southern hemisphere like 
SGSO may help us to gain insight into the diffusion pattern around B1055-52 and 
make a certain judgment to the relation between B1055-52 and the positron 
excess.

\acknowledgments{This work is supported by the National Key Program for 
Research and Development (No.~2016YFA0400200) and by the National Natural 
Science Foundation of China under Grants No.~U1738209,~11851303.}

\bibliography{positron}

\begin{thebibliography}{}
\expandafter\ifx\csname natexlab\endcsname\relax\def\natexlab#1{#1}\fi
\providecommand{\url}[1]{\href{#1}{#1}}

\bibitem[{{Abdo} {et~al.}(2010){Abdo}, {Ackermann}, {Ajello}, {Atwood},
  {Axelsson}, {Baldini}, {Ballet}, {Barbiellini}, {Baring}, {Bastieri},
  {Baughman}, {Bechtol}, {Bellazzini}, {Berenji}, {Blandford}, {Bloom},
  {Bonamente}, {Borgland}, {Bregeon}, {Brez}, {Brigida}, {Bruel}, {Burnett},
  {Buson}, {Caliandro}, {Cameron}, {Camilo}, {Caraveo}, {Casandjian}, {Cecchi},
  {{\c{C}}elik}, {Charles}, {Chekhtman}, {Cheung}, {Chiang}, {Ciprini},
  {Claus}, {Cognard}, {Cohen-Tanugi}, {Cominsky}, {Conrad}, {Corbet}, {Cutini},
  {den Hartog}, {Dermer}, {de Angelis}, {de Luca}, {de Palma}, {Digel},
  {Dormody}, {Silva}, {Drell}, {Dubois}, {Dumora}, {Espinoza}, {Farnier},
  {Favuzzi}, {Fegan}, {Ferrara}, {Focke}, {Fortin}, {Frailis}, {Freire},
  {Fukazawa}, {Funk}, {Fusco}, {Gargano}, {Gasparrini}, {Gehrels}, {Germani},
  {Giavitto}, {Giebels}, {Giglietto}, {Giommi}, {Giordano}, {Glanzman},
  {Godfrey}, {Gotthelf}, {Grenier}, {Grondin}, {Grove}, {Guillemot}, {Guiriec},
  {Gwon}, {Hanabata}, {Harding}, {Hayashida}, {Hays}, {Hughes}, {Jackson},
  {J{\'o}hannesson}, {Johnson}, {Johnson}, {Johnson}, {Johnson}, {Johnston},
  {Kamae}, {Kanbach}, {Kaspi}, {Katagiri}, {Kataoka}, {Kawai}, {Kerr},
  {Kn{\"o}dlseder}, {Kocian}, {Kramer}, {Kuss}, {Lande}, {Latronico},
  {Lemoine-Goumard}, {Livingstone}, {Longo}, {Loparco}, {Lott}, {Lovellette},
  {Lubrano}, {Lyne}, {Madejski}, {Makeev}, {Manchester}, {Marelli},
  {Mazziotta}, {McConville}, {McEnery}, {McGlynn}, {Meurer}, {Michelson},
  {Mineo}, {Mitthumsiri}, {Mizuno}, {Moiseev}, {Monte}, {Monzani}, {Morselli},
  {Moskalenko}, {Murgia}, {Nakamori}, {Nolan}, {Norris}, {Noutsos}, {Nuss},
  {Ohsugi}, {Omodei}, {Orlando}, {Ormes}, {Ozaki}, {Paneque}, {Panetta},
  {Parent}, {Pelassa}, {Pepe}, {Pesce-Rollins}, {Piron}, {Porter}, {Rain{\`o}},
  {Rando}, {Ransom}, {Ray}, {Razzano}, {Rea}, {Reimer}, {Reimer}, {Reposeur},
  {Ritz}, {Rodriguez}, {Romani}, {Roth}, {Ryde}, {Sadrozinski}, {Sanchez},
  {Sander}, {Saz Parkinson}, {Scargle}, {Schalk}, {Sellerholm}, {Sgr{\`o}},
  {Siskind}, {Smith}, {Smith}, {Spandre}, {Spinelli}, {Stappers}, {Starck},
  {Striani}, {Strickman}, {Strong}, {Suson}, {Tajima}, {Takahashi},
  {Takahashi}, {Tanaka}, {Thayer}, {Thayer}, {Theureau}, {Thompson},
  {Thorsett}, {Tibaldo}, {Tibolla}, {Torres}, {Tosti}, {Tramacere}, {Uchiyama},
  {Usher}, {Van Etten}, {Vasileiou}, {Venter}, {Vilchez}, {Vitale}, {Waite},
  {Wang}, {Wang}, {Watters}, {Weltevrede}, {Winer}, {Wood}, {Ylinen}, \&
  {Ziegler}}]{2010ApJS..187..460A}
{Abdo}, A.~A., {Ackermann}, M., {Ajello}, M., {et~al.} 2010, \apjs, 187, 460

\bibitem[{{Abeysekara} {et~al.}(2017{\natexlab{a}}){Abeysekara}, {Albert},
  {Alfaro}, {et~al.}}]{2017Sci...358..911A}
{Abeysekara}, A.~U., {Albert}, A., {Alfaro}, R., {et~al.} 2017{\natexlab{a}},
  Science, 358, 911

\bibitem[{{Abeysekara} {et~al.}(2017{\natexlab{b}}){Abeysekara}, {Albert},
  {Alfaro}, {Alvarez}, {{\'A}lvarez}, {Arceo}, {Arteaga-Vel{\'a}zquez}, {Ayala
  Solares}, {Barber}, {Baughman}, {Bautista-Elivar}, {Becerra Gonzalez},
  {Becerril}, {Belmont-Moreno}, {BenZvi}, {Berley}, {Bernal}, {Braun},
  {Brisbois}, {Caballero-Mora}, {Capistr{\'a}n}, {Carrami{\~n}ana}, {Casanova},
  {Castillo}, {Cotti}, {Cotzomi}, {Couti{\~n}o de Le{\'o}n}, {de la Fuente},
  {De Le{\'o}n}, {Diaz Hernandez}, {Dingus}, {DuVernois},
  {D{\'{\i}}az-V{\'e}lez}, {Ellsworth}, {Engel}, {Fiorino}, {Fraija},
  {Garc{\'{\i}}a-Gonz{\'a}lez}, {Garfias}, {Gerhardt}, {Gonz{\'a}lez
  Mu{\~n}oz}, {Gonz{\'a}lez}, {Goodman}, {Hampel-Arias}, {Harding},
  {Hernandez}, {Hernandez-Almada}, {Hinton}, {Hui}, {H{\"u}ntemeyer},
  {Iriarte}, {Jardin-Blicq}, {Joshi}, {Kaufmann}, {Kieda}, {Lara}, {Lauer},
  {Lee}, {Lennarz}, {Le{\'o}n Vargas}, {Linnemann}, {Longinotti}, {Raya},
  {Luna-Garc{\'{\i}}a}, {L{\'o}pez-Coto}, {Malone}, {Marinelli}, {Martinez},
  {Martinez-Castellanos}, {Mart{\'{\i}}nez-Castro}, {Mart{\'{\i}}nez-Huerta},
  {Matthews}, {Miranda-Romagnoli}, {Moreno}, {Mostaf{\'a}}, {Nellen},
  {Newbold}, {Nisa}, {Noriega-Papaqui}, {Pelayo}, {Pretz},
  {P{\'e}rez-P{\'e}rez}, {Ren}, {Rho}, {Rivi{\`e}re}, {Rosa-Gonz{\'a}lez},
  {Rosenberg}, {Ruiz-Velasco}, {Salazar}, {Salesa Greus}, {Sandoval},
  {Schneider}, {Schoorlemmer}, {Sinnis}, {Smith}, {Springer}, {Surajbali},
  {Taboada}, {Tibolla}, {Tollefson}, {Torres}, {Ukwatta}, {Vianello},
  {Villase{\~n}or}, {Weisgarber}, {Westerhoff}, {Wisher}, {Wood}, {Yapici},
  {Younk}, {Zepeda}, \& {Zhou}}]{2017ApJ...843...40A}
---. 2017{\natexlab{b}}, \apj, 843, 40

\bibitem[{{Ackermann} {et~al.}(2012){Ackermann}, {Ajello}, {Allafort},
  {Atwood}, {Baldini}, {Barbiellini}, {Bastieri}, {Bechtol}, {Bellazzini},
  {Berenji}, {Blandford}, {Bloom}, {Bonamente}, {Borgland}, {Bouvier},
  {Bregeon}, {Brigida}, {Bruel}, {Buehler}, {Buson}, {Caliandro}, {Cameron},
  {Caraveo}, {Casandjian}, {Cecchi}, {Charles}, {Chekhtman}, {Cheung},
  {Chiang}, {Ciprini}, {Claus}, {Cohen-Tanugi}, {Conrad}, {Cutini}, {de
  Angelis}, {de Palma}, {Dermer}, {Digel}, {Do Couto E Silva}, {Drell},
  {Drlica-Wagner}, {Favuzzi}, {Fegan}, {Ferrara}, {Focke}, {Fortin},
  {Fukazawa}, {Funk}, {Fusco}, {Gargano}, {Gasparrini}, {Germani}, {Giglietto},
  {Giommi}, {Giordano}, {Giroletti}, {Glanzman}, {Godfrey}, {Grenier}, {Grove},
  {Guiriec}, {Gustafsson}, {Hadasch}, {Harding}, {Hayashida}, {Hughes},
  {J{\'o}hannesson}, {Johnson}, {Kamae}, {Katagiri}, {Kataoka},
  {Kn{\"o}dlseder}, {Kuss}, {Lande}, {Latronico}, {Lemoine-Goumard}, {Llena
  Garde}, {Longo}, {Loparco}, {Lovellette}, {Lubrano}, {Madejski}, {Mazziotta},
  {McEnery}, {Michelson}, {Mitthumsiri}, {Mizuno}, {Moiseev}, {Monte},
  {Monzani}, {Morselli}, {Moskalenko}, {Murgia}, {Nakamori}, {Nolan}, {Norris},
  {Nuss}, {Ohno}, {Ohsugi}, {Okumura}, {Omodei}, {Orlando}, {Ormes}, {Ozaki},
  {Paneque}, {Parent}, {Pesce-Rollins}, {Pierbattista}, {Piron}, {Pivato},
  {Porter}, {Rain{\`o}}, {Rando}, {Razzano}, {Razzaque}, {Reimer}, {Reimer},
  {Reposeur}, {Ritz}, {Romani}, {Roth}, {Sadrozinski}, {Sbarra}, {Schalk},
  {Sgr{\`o}}, {Siskind}, {Spandre}, {Spinelli}, {Strong}, {Takahashi},
  {Takahashi}, {Tanaka}, {Thayer}, {Thayer}, {Tibaldo}, {Tinivella}, {Torres},
  {Tosti}, {Troja}, {Uchiyama}, {Usher}, {Vandenbroucke}, {Vasileiou},
  {Vianello}, {Vitale}, {Waite}, {Winer}, {Wood}, {Wood}, {Yang}, \&
  {Zimmer}}]{2012PhRvL.108a1103A}
{Ackermann}, M., {Ajello}, M., {Allafort}, A., {et~al.} 2012, Physical Review
  Letters, 108, 011103

\bibitem[{{Ackermann} {et~al.}(2015){Ackermann}, {Albert}, {Anderson},
  {Atwood}, {Baldini}, {Barbiellini}, {Bastieri}, {Bechtol}, {Bellazzini}, \&
  {Bissaldi}}]{2015PhRvL.115w1301A}
{Ackermann}, M., {Albert}, A., {Anderson}, B., {et~al.} 2015, \prl, 115, 231301

\bibitem[{{Adriani} {et~al.}(2009){Adriani}, {Barbarino}, {Bazilevskaya},
  {Bellotti}, {Boezio}, {Bogomolov}, {Bonechi}, {Bongi}, {Bonvicini}, {Bottai},
  {Bruno}, {Cafagna}, {Campana}, {Carlson}, {Casolino}, {Castellini}, {de
  Pascale}, {de Rosa}, {de Simone}, {di Felice}, {Galper}, {Grishantseva},
  {Hofverberg}, {Koldashov}, {Krutkov}, {Kvashnin}, {Leonov}, {Malvezzi},
  {Marcelli}, {Menn}, {Mikhailov}, {Mocchiutti}, {Orsi}, {Osteria}, {Papini},
  {Pearce}, {Picozza}, {Ricci}, {Ricciarini}, {Simon}, {Sparvoli},
  {Spillantini}, {Stozhkov}, {Vacchi}, {Vannuccini}, {Vasilyev}, {Voronov},
  {Yurkin}, {Zampa}, {Zampa}, \& {Zverev}}]{2009Natur.458..607A}
{Adriani}, O., {Barbarino}, G.~C., {Bazilevskaya}, G.~A., {et~al.} 2009, \nat,
  458, 607

\bibitem[{{Aguilar} {et~al.}(2013){Aguilar}, {Alberti}, {Alpat}, {Alvino},
  {Ambrosi}, {Andeen}, {Anderhub}, {Arruda}, {Azzarello}, {Bachlechner}, \&
  et~al.}]{2013PhRvL.110n1102A}
{Aguilar}, M., {Alberti}, G., {Alpat}, B., {et~al.} 2013, Physical Review
  Letters, 110, 141102

\bibitem[{{Aguilar} {et~al.}(2016){Aguilar}, {Ali Cavasonza}, {Ambrosi},
  {Arruda}, {Attig}, {Aupetit}, {Azzarello}, {Bachlechner}, {Barao}, {Barrau},
  \& et~al.}]{2016PhRvL.117w1102A}
{Aguilar}, M., {Ali Cavasonza}, L., {Ambrosi}, G., {et~al.} 2016, Physical
  Review Letters, 117, 231102

\bibitem[{{Aguilar} {et~al.}(2019){Aguilar}, {Ali Cavasonza}, {Ambrosi},
  {Arruda}, {Attig}, {Azzarello}, {Bachlechner}, {Barao}, {Barrau}, {Barrin},
  \& et~al.}]{2019PhRvL.122d1102A}
---. 2019, Physical Review Letters, 122, 041102

\bibitem[{{Albert} {et~al.}(2019){Albert}, {Alfaro}, {Ashkar}, {Alvarez},
  {{\'A}lvarez}, {Arteaga-Vel{\'a}zquez}, {Ayala Solares}, {Arceo}, {Bellido},
  \& {BenZvi}}]{2019arXiv190208429A}
{Albert}, A., {Alfaro}, R., {Ashkar}, H., {et~al.} 2019, arXiv e-prints,
  arXiv:1902.08429

\bibitem[{{Aliu} {et~al.}(2014){Aliu}, {Archambault}, {Aune}, {Behera},
  {Beilicke}, {Benbow}, {Berger}, {Bird}, {Buckley}, {Bugaev}, {Cardenzana},
  {Cerruti}, {Chen}, {Ciupik}, {Collins-Hughes}, {Connolly}, {Cui}, {Dumm},
  {Dwarkadas}, {Errando}, {Falcone}, {Federici}, {Feng}, {Finley},
  {Fleischhack}, {Fortin}, {Fortson}, {Furniss}, {Galante}, {Gall},
  {Gillanders}, {Griffin}, {Griffiths}, {Grube}, {Gyuk}, {Hanna}, {Holder},
  {Hughes}, {Humensky}, {Kaaret}, {Kertzman}, {Khassen}, {Kieda}, {Krennrich},
  {Kumar}, {Lang}, {Madhavan}, {Maier}, {McCann}, {Meagher}, {Millis},
  {Moriarty}, {Mukherjee}, {Nieto}, {O'Faol{\'a}in de Bhr{\'o}ithe}, {Ong},
  {Otte}, {Pandel}, {Park}, {Pohl}, {Popkow}, {Prokoph}, {Quinn}, {Ragan},
  {Rajotte}, {Ratliff}, {Reyes}, {Reynolds}, {Richards}, {Roache}, {Rousselle},
  {Sembroski}, {Shahinyan}, {Sheidaei}, {Smith}, {Staszak}, {Telezhinsky},
  {Tsurusaki}, {Tucci}, {Tyler}, {Varlotta}, {Vassiliev}, {Vincent}, {Wakely},
  {Ward}, {Weinstein}, {Welsing}, \& {Wilhelm}}]{2014ApJ...787..166A}
{Aliu}, E., {Archambault}, S., {Aune}, T., {et~al.} 2014, \apj, 787, 166

\bibitem[{{Becker} \& {Truemper}(1997)}]{1997A&A...326..682B}
{Becker}, W., \& {Truemper}, J. 1997, \aap, 326, 682

\bibitem[{{B{\^\i}rzan} {et~al.}(2016){B{\^\i}rzan}, {Pavlov}, \&
  {Kargaltsev}}]{2016ApJ...817..129B}
{B{\^\i}rzan}, L., {Pavlov}, G.~G., \& {Kargaltsev}, O. 2016, \apj, 817, 129

\bibitem[{{Brisken} {et~al.}(2003){Brisken}, {Thorsett}, {Golden}, \&
  {Goss}}]{2003ApJ...593L..89B}
{Brisken}, W.~F., {Thorsett}, S.~E., {Golden}, A., \& {Goss}, W.~M. 2003,
  \apjl, 593, L89

\bibitem[{{Caraveo} {et~al.}(2003){Caraveo}, {Bignami}, {De Luca},
  {Mereghetti}, {Pellizzoni}, {Mignani}, {Tur}, \&
  {Becker}}]{2003Sci...301.1345C}
{Caraveo}, P.~A., {Bignami}, G.~F., {De Luca}, A., {et~al.} 2003, Science, 301,
  1345

\bibitem[{{Cordes} \& {Lazio}(2002)}]{2002astro.ph..7156C}
{Cordes}, J.~M., \& {Lazio}, T.~J.~W. 2002, arXiv Astrophysics e-prints,
  astro-ph/0207156

\bibitem[{{Delahaye} {et~al.}(2010){Delahaye}, {Lavalle}, {Lineros}, {Donato},
  \& {Fornengo}}]{dela10}
{Delahaye}, T., {Lavalle}, J., {Lineros}, R., {Donato}, F., \& {Fornengo}, N.
  2010, \aap, 524, A51

\bibitem[{{Della Torre} {et~al.}(2015){Della Torre}, {Gervasi}, {Rancoita},
  {Rozza}, \& {Treves}}]{2015JHEAp...8...27D}
{Della Torre}, S., {Gervasi}, M., {Rancoita}, P.~G., {Rozza}, D., \& {Treves},
  A. 2015, Journal of High Energy Astrophysics, 8, 27

\bibitem[{{Di Mauro} {et~al.}(2019){Di Mauro}, {Manconi}, \&
  {Donato}}]{2019arXiv190305647D}
{Di Mauro}, M., {Manconi}, S., \& {Donato}, F. 2019, arXiv e-prints,
  arXiv:1903.05647

\bibitem[{{Evoli} {et~al.}(2018){Evoli}, {Linden}, \&
  {Morlino}}]{2018PhRvD..98f3017E}
{Evoli}, C., {Linden}, T., \& {Morlino}, G. 2018, \prd, 98, 063017

\bibitem[{{Fang} {et~al.}(2019){Fang}, {Bi}, \& {Yin}}]{2019arXiv190306421F}
{Fang}, K., {Bi}, X.-J., \& {Yin}, P.-F. 2019, arXiv e-prints, arXiv:1903.06421

\bibitem[{{Fang} {et~al.}(2018){Fang}, {Bi}, {Yin}, \&
  {Yuan}}]{2018ApJ...863...30F}
{Fang}, K., {Bi}, X.-J., {Yin}, P.-F., \& {Yuan}, Q. 2018, \apj, 863, 30

\bibitem[{{H.~E.~S.~S. Collaboration} {et~al.}(2018){H.~E.~S.~S.
  Collaboration}, {Abdalla}, {Abramowski}, {Aharonian}, {Ait Benkhali},
  {Ang{\"u}ner}, {Arakawa}, {Arrieta}, {Aubert}, {Backes}, {Balzer}, {Barnard},
  {Becherini}, {Becker Tjus}, {Berge}, {Bernhard}, {Bernl{\"o}hr}, {Blackwell},
  {B{\"o}ttcher}, {Boisson}, {Bolmont}, {Bonnefoy}, {Bordas}, {Bregeon},
  {Brun}, {Brun}, {Bryan}, {B{\"u}chele}, {Bulik}, {Capasso}, {Carrigan},
  {Caroff}, {Carosi}, {Casanova}, {Cerruti}, {Chakraborty}, {Chaves}, {Chen},
  {Chevalier}, {Colafrancesco}, {Condon}, {Conrad}, {Davids}, {Decock}, {Deil},
  {Devin}, {deWilt}, {Dirson}, {Djannati-Ata{\"\i}}, {Domainko}, {Donath},
  {Drury}, {Dutson}, {Dyks}, {Edwards}, {Egberts}, {Eger}, {Emery},
  {Ernenwein}, {Eschbach}, {Farnier}, {Fegan}, {Fernand es}, {Fiasson},
  {Fontaine}, {F{\"o}rster}, {Funk}, {F{\"u}{\ss}ling}, {Gabici}, {Gallant},
  {Garrigoux}, {Gast}, {Gat{\'e}}, {Giavitto}, {Giebels}, {Glawion},
  {Glicenstein}, {Gottschall}, {Grondin}, {Hahn}, {Haupt}, {Hawkes},
  {Heinzelmann}, {Henri}, {Hermann}, {Hinton}, {Hofmann}, {Hoischen}, {Holch},
  {Holler}, {Horns}, {Ivascenko}, {Iwasaki}, {Jacholkowska}, {Jamrozy},
  {Jankowsky}, {Jankowsky}, {Jingo}, {Jouvin}, {Jung-Richardt}, {Kastendieck},
  {Katarzy{\'n}ski}, {Katsuragawa}, {Katz}, {Kerszberg}, {Khangulyan},
  {Kh{\'e}lifi}, {King}, {Klepser}, {Klochkov}, {Klu{\'z}niak}, {Komin},
  {Kosack}, {Krakau}, {Kraus}, {Kr{\"u}ger}, {Laffon}, {Lamanna}, {Lau},
  {Lees}, {Lefaucheur}, {Lemi{\`e}re}, {Lemoine-Goumard}, {Lenain}, {Leser},
  {Lohse}, {Lorentz}, {Liu}, {L{\'o}pez-Coto}, {Lypova}, {Marandon},
  {Malyshev}, {Marcowith}, {Mariaud}, {Marx}, {Maurin}, {Maxted}, {Mayer},
  {Meintjes}, {Meyer}, {Mitchell}, {Moderski}, {Mohamed}, {Mohrmann},
  {Mor{\r{a}}}, {Moulin}, {Murach}, {Nakashima}, {de Naurois}, {Ndiyavala},
  {Niederwanger}, {Niemiec}, {Oakes}, {O'Brien}, {Odaka}, {Ohm}, {Ostrowski},
  {Oya}, {Padovani}, {Panter}, {Parsons}, {Paz Arribas}, {Pekeur}, {Pelletier},
  {Perennes}, {Petrucci}, {Peyaud}, {Piel}, {Pita}, {Poireau}, {Poon},
  {Prokhorov}, {Prokoph}, {P{\"u}hlhofer}, {Punch}, {Quirrenbach}, {Raab},
  {Rauth}, {Reimer}, {Reimer}, {Renaud}, {de los Reyes}, {Rieger}, {Rinchiuso},
  {Romoli}, {Rowell}, {Rudak}, {Rulten}, {Safi-Harb}, {Sahakian}, {Saito},
  {Sanchez}, {Santangelo}, {Sasaki}, {Schand ri}, {Schlickeiser},
  {Sch{\"u}ssler}, {Schulz}, {Schwanke}, {Schwemmer}, {Seglar-Arroyo},
  {Settimo}, {Seyffert}, {Shafi}, {Shilon}, {Shiningayamwe}, {Simoni}, {Sol},
  {Spanier}, {Spir-Jacob}, {Stawarz}, {Steenkamp}, {Stegmann}, {Steppa},
  {Sushch}, {Takahashi}, {Tavernet}, {Tavernier}, {Taylor}, {Terrier},
  {Tibaldo}, {Tiziani}, {Tluczykont}, {Trichard}, {Tsirou}, {Tsuji}, {Tuffs},
  {Uchiyama}, {van der Walt}, {van Eldik}, {van Rensburg}, {van Soelen},
  {Vasileiadis}, {Veh}, {Venter}, {Viana}, {Vincent}, {Vink}, {Voisin},
  {V{\"o}lk}, {Vuillaume}, {Wadiasingh}, {Wagner}, {Wagner}, {Wagner}, {White},
  {Wierzcholska}, {Willmann}, {W{\"o}rnlein}, {Wouters}, {Yang}, {Zaborov},
  {Zacharias}, {Zanin}, {Zdziarski}, {Zech}, {Zefi}, {Ziegler}, {Zorn}, \&
  {{\.Z}ywucka}}]{2018A&A...612A...1H}
{H.~E.~S.~S. Collaboration}, {Abdalla}, H., {Abramowski}, A., {et~al.} 2018,
  \aap, 612, A1

\bibitem[{{Hooper} {et~al.}(2009){Hooper}, {Blasi}, \& {Dario
  Serpico}}]{2009JCAP...01..025H}
{Hooper}, D., {Blasi}, P., \& {Dario Serpico}, P. 2009, \jcap, 1, 25

\bibitem[{{Hooper} {et~al.}(2017){Hooper}, {Cholis}, {Linden}, \&
  {Fang}}]{2017PhRvD..96j3013H}
{Hooper}, D., {Cholis}, I., {Linden}, T., \& {Fang}, K. 2017, \prd, 96, 103013

\bibitem[{{Huang} {et~al.}(2018){Huang}, {Fang}, {Liu}, \&
  {Wang}}]{2018ApJ...866..143H}
{Huang}, Z.-Q., {Fang}, K., {Liu}, R.-Y., \& {Wang}, X.-Y. 2018, \apj, 866, 143

\bibitem[{{Khangulyan} {et~al.}(2018){Khangulyan}, {Koldoba}, {Ustyugova},
  {Bogovalov}, \& {Aharonian}}]{2018ApJ...860...59K}
{Khangulyan}, D., {Koldoba}, A.~V., {Ustyugova}, G.~V., {Bogovalov}, S.~V., \&
  {Aharonian}, F. 2018, \apj, 860, 59

\bibitem[{{Kirk} \& {Duffy}(1999)}]{1999JPhG...25R.163K}
{Kirk}, J.~G., \& {Duffy}, P. 1999, Journal of Physics G Nuclear Physics, 25,
  R163

\bibitem[{{Lin} {et~al.}(2015){Lin}, {Yuan}, \& {Bi}}]{2015PhRvD..91f3508L}
{Lin}, S.-J., {Yuan}, Q., \& {Bi}, X.-J. 2015, \prd, 91, 063508

\bibitem[{{Linden} {et~al.}(2017){Linden}, {Auchettl}, {Bramante}, {Cholis},
  {Fang}, {Hooper}, {Karwal}, \& {Li}}]{2017PhRvD..96j3016L}
{Linden}, T., {Auchettl}, K., {Bramante}, J., {et~al.} 2017, \prd, 96, 103016

\bibitem[{{Liu} {et~al.}(2019){Liu}, {Yan}, \& {Zhang}}]{2019arXiv190411536L}
{Liu}, R.-Y., {Yan}, H., \& {Zhang}, H. 2019, arXiv e-prints, arXiv:1904.11536

\bibitem[{{Liu} {et~al.}(2017){Liu}, {Bi}, {Lin}, \&
  {Yin}}]{2017ChPhC..41d5104L}
{Liu}, W., {Bi}, X.-J., {Lin}, S.-J., \& {Yin}, P.-F. 2017, Chinese Physics C,
  41, 045104

\bibitem[{{L{\'o}pez-Coto} \& {Giacinti}(2018)}]{2018MNRAS.479.4526L}
{L{\'o}pez-Coto}, R., \& {Giacinti}, G. 2018, \mnras, 479, 4526

\bibitem[{{Manchester} {et~al.}(2005){Manchester}, {Hobbs}, {Teoh}, \&
  {Hobbs}}]{2005AJ....129.1993M}
{Manchester}, R.~N., {Hobbs}, G.~B., {Teoh}, A., \& {Hobbs}, M. 2005, \aj, 129,
  1993

\bibitem[{{Mignani} {et~al.}(2010){Mignani}, {Pavlov}, \&
  {Kargaltsev}}]{2010ApJ...720.1635M}
{Mignani}, R.~P., {Pavlov}, G.~G., \& {Kargaltsev}, O. 2010, \apj, 720, 1635

\bibitem[{{Minter} \& {Spangler}(1996)}]{1996ApJ...458..194M}
{Minter}, A.~H., \& {Spangler}, S.~R. 1996, \apj, 458, 194

\bibitem[{{Muslimov} \& {Harding}(2003)}]{2003ApJ...588..430M}
{Muslimov}, A.~G., \& {Harding}, A.~K. 2003, \apj, 588, 430

\bibitem[{{Pacini} \& {Salvati}(1973)}]{1973ApJ...186..249P}
{Pacini}, F., \& {Salvati}, M. 1973, \apj, 186, 249

\bibitem[{{Posselt} {et~al.}(2015){Posselt}, {Spence}, \&
  {Pavlov}}]{2015ApJ...811...96P}
{Posselt}, B., {Spence}, G., \& {Pavlov}, G.~G. 2015, \apj, 811, 96

\bibitem[{{Posselt} {et~al.}(2017){Posselt}, {Pavlov}, {Slane}, {Romani},
  {Bucciantini}, {Bykov}, {Kargaltsev}, {Weisskopf}, \&
  {Ng}}]{2017ApJ...835...66P}
{Posselt}, B., {Pavlov}, G.~G., {Slane}, P.~O., {et~al.} 2017, \apj, 835, 66

\bibitem[{{Profumo} {et~al.}(2018){Profumo}, {Reynoso-Cordova}, {Kaaz}, \&
  {Silverman}}]{2018PhRvD..97l3008P}
{Profumo}, S., {Reynoso-Cordova}, J., {Kaaz}, N., \& {Silverman}, M. 2018,
  \prd, 97, 123008

\bibitem[{{Reynolds} {et~al.}(2017){Reynolds}, {Pavlov}, {Kargaltsev},
  {Klingler}, {Renaud}, \& {Mereghetti}}]{2017SSRv..207..175R}
{Reynolds}, S.~P., {Pavlov}, G.~G., {Kargaltsev}, O., {et~al.} 2017, \ssr, 207,
  175

\bibitem[{{Schlickeiser} \& {Ruppel}(2010)}]{schli10}
{Schlickeiser}, R., \& {Ruppel}, J. 2010, New Journal of Physics, 12, 033044

\bibitem[{{Tang} \& {Piran}(2019)}]{2019MNRAS.484.3491T}
{Tang}, X., \& {Piran}, T. 2019, \mnras, 484, 3491

\bibitem[{{Taylor} \& {Cordes}(1993)}]{1993ApJ...411..674T}
{Taylor}, J.~H., \& {Cordes}, J.~M. 1993, \apj, 411, 674

\bibitem[{{Xi} {et~al.}(2018){Xi}, {Liu}, {Huang}, {Fang}, {Yan}, \&
  {Wang}}]{2018arXiv181010928X}
{Xi}, S.-Q., {Liu}, R.-Y., {Huang}, Z.-Q., {et~al.} 2018, arXiv e-prints,
  arXiv:1810.10928

\bibitem[{{Yao} {et~al.}(2017){Yao}, {Manchester}, \&
  {Wang}}]{2017ApJ...835...29Y}
{Yao}, J.~M., {Manchester}, R.~N., \& {Wang}, N. 2017, \apj, 835, 29

\bibitem[{{Yin} {et~al.}(2013){Yin}, {Yu}, {Yuan}, \&
  {Bi}}]{2013PhRvD..88b3001Y}
{Yin}, P.-F., {Yu}, Z.-H., {Yuan}, Q., \& {Bi}, X.-J. 2013, \prd, 88, 023001

\bibitem[{{Yuan} {et~al.}(2017){Yuan}, {Lin}, {Fang}, \&
  {Bi}}]{2017PhRvD..95h3007Y}
{Yuan}, Q., {Lin}, S.-J., {Fang}, K., \& {Bi}, X.-J. 2017, \prd, 95, 083007

\bibitem[{{Y{\"u}ksel} {et~al.}(2009){Y{\"u}ksel}, {Kistler}, \&
  {Stanev}}]{2009PhRvL.103e1101Y}
{Y{\"u}ksel}, H., {Kistler}, M.~D., \& {Stanev}, T. 2009, Physical Review
  Letters, 103, 051101

\bibitem[{{Zhang} {et~al.}(2004){Zhang}, {Cheng}, {Jiang}, \&
  {Leung}}]{2004ApJ...604..317Z}
{Zhang}, L., {Cheng}, K.~S., {Jiang}, Z.~J., \& {Leung}, P. 2004, \apj, 604,
  317

\end{thebibliography}

\end{document}